# Nonlinear Dynamics, Lorenz Model and Formation of Binary Stars

Yi-Fang Chang

*Department of Physics, Yunnan University, Kunming, 650091, China*

(E-mail: yifangchang1030@hotmail.com)

**ABSTRACT**

Based on the Lorenz model derived from the equations of hydrodynamics of nebula, we discuss the formation of binary stars by the qualitative analysis theory of nonlinear equation. Here the two wings in the Lorenz model form just the binary stars, whose Roche surface is result of evolution under certain condition. The nonlinear interaction plays a crucial role, and is necessary condition of the formation of binary stars and of multiple stars. While the linear equations form only a single star. Moreover, the Lorenz model may be obtained from synergetics yet, and some dynamical equations of the order parameter can also form the binary star system.

Now the common existence of the binary star systems and the explanation are notable problems in astronomy [1-7]. Itoh, et al., discussed the equation of motion for relativistic compact binaries with the strong field [8]. Ford, et al., derived octupole-level perturbation equation for hierarchical triple star systems, which describes the secular evolution of the orbital eccentricities and inclinations [9]. One of more successful theories of the formation of binary stars is the fragmentation proposed by Boss, et al. [10-12], which supposes that binary stars are born during the protostellar collapse phase under their own gravity. Using computer simulation they obtained that an initial spherical cloud in rapidly rotation collapses and flattens to a disk, which later fragments into a binary system.

Based on the basic equations of a rotating disk on the nebula, we applied the qualitative analysis theory of nonlinear equation, and obtained a nonlinear dynamical model of formation of binary stars [13]. Under certain conditions a pair of singular points results in the course of evolution, which corresponds to the binary stars. Under other conditions these equations give a single central point, which corresponds to a single star. This method and model may be extended and developed. Steinitz and Farbiash established the correlation between the spins (rotational velocities) in binaries, and show that the degree of spin correlation is independent of the components' separation. Such a result might be related for example to Zhang's nonlinear model for the formation of binary stars from a nebula [14]. The general relativity is the base for any exact evolutionary theory of large scale structures. We calculate the universal 2+1-dimensional plane equations of gravitational field in general relativity. Based on these equations, the evolution of disk nebula is discussed. A system of nebula can form binary stars or single star for different conditions. While any simplified linear theory forms only a single star system. It is proved that the nonlinear interactions are very general, so the binary stars are also common [15]. Further, based on the hydrodynamics and hydromagnetics of nebula, from Alfver equation of the cosmical electrodynamics [16] we discussed the formation of binary stars by the qualitative analysis theory of nonlinear equation [17].

Mardling discussed the role of chaos in the circularization of tidal capture binary stars [4]. But, this quantitative process is a problem still. In this paper, we start from the Lorenz model and other nonlinear theory, and prove that the nonlinear interactions are necessary condition of the formation of binary stars.

The base of our model is that a star is formed from nebula. While nebula, in which there is mainly hydrogen and helium plasma, obeys the nonlinear hydromagnetic equations. Their forms are [16]

$$\rho \frac{du}{dt} = \rho[\frac{\partial u}{\partial t} + (u\nabla)u] = G + \frac{e}{c}u \times B - gradp + v\Delta u + \frac{\eta}{3} graddivu. \quad (1)$$

When the magnetic field and G are neglected, and $\eta = 0$, Eq. (1) of the two dimensions combines the equation of continuity with the convection:

$$\rho_t + u\rho_x + w\rho_y = (N^2/g)w + k\Delta\rho. \quad (2)$$

Then the Saltzman model is derived [18]. From this the well-known Lorenz model may be derived [19]. Its equations are:

$$dx/dt = -vx + ky, \quad (3)$$

$$dy/dt = ax - by - xz, \quad (4)$$

$$dz/dt = -cz + xy, \quad (5)$$

where $x$ is the flow rate, $y$ and $z$ are asymmetric and symmetric parts of temperature difference on nebula, respectively. These are a simplified result of the Navier-Stokes equations. Usually, we suppose that all parameters are positive. If various parameters in Eqs.(3)(4)(5) take suitable values, a beautiful Lorenz strange attractor will be obtained. The attractor possesses some similarities to the final synthesis map, which shows that bipolar jets emerge from a central core [20]. The astronomical meaning of the Lorenz model is that a nebula passes through chaos, and becomes a pair of the strange attractor, which may form binary stars due to self-gravitation.

Using a chaotic mechanism of the Lorenz model we discuss the formation of a binary star system from two-dimensional nebula.

1. If $dx/dt = 0$, $x = (k/v)y = ey$. Eqs.(4) and (5) become:

$$dy/dt = (ae - b)y - eyz, \quad (6)$$

$$dz/dt = -cz + ey^2, \quad . \quad (7)$$

Using the qualitative analysis theory of the nonlinear equations, the characteristic matrix of Eqs.(6) and (7) is

$$\begin{pmatrix} ae - b - ez & -ey \\ 2ey & -c \end{pmatrix}. \quad (8)$$

Its characteristic equation is

$$\lambda^2 - T\lambda + D = \lambda^2 - (ae - b - ez - c)\lambda - c(ae - b - ez) + 2e^2 y^2 = 0. \quad (9)$$

Here a plane system of $dy/dt = 0$ and $dz/dt = 0$ has three singular (equilibrium) points: $O(0,0)$, $A$ and $B(\pm\sqrt{c(ae-b)}/e, a-(b/e))$. For $O$ point, $\lambda_{1,2} = ae-b, -c$. When $ae > 0$, it is a saddle point; when $ae < 0$, it is a nodal point. For $A$ and $B$ points, $T = -c < 0$, so they are two stable sinks, and $\lambda_{1,2} = [-c \pm \sqrt{c(c-8ae+8b)}]/2$. When $8ae > c+8b$, they are two focal points; when $8ae < c+8b$, they are two nodal points. The system possesses the saddle-focal points (for $8ae > 8b+c > 8b$), the saddle-nodal points (for $8b+c > 8ae > 8b$) and the nodal-nodal points (for $8b+c > 8b > 8ae$), respectively. In these cases, the binary stars may be formed. These saddle points are some critical points, which separate different attractive basins; while these basins form different stars.

As a particular case, if $ae - b - ez - c = 0$, so $z = (ae-b-c)/e$. Eq.(6) becomes:

$$dy/dt = cy. \qquad (10)$$

So it is obtained that $y_{1,2} = \pm\sqrt{c(ae-b-c)}/e$ from (7), and $\lambda_{1,2} = \pm\sqrt{c(3c+2b-2ae)}$ from (9). When $3c+2b > 2ae$, $A$ and $B$ are two saddle points; when $3c+2b < 2ae$, $A$ and $B$ are two central points, which produces a Roche surface of binary stars.

2. If $dy/dt = 0$, $y = x(a-z)/b$. Eqs.(3) and (5) become:

$$dx/dt = [-v + (ka/b)]x - (k/b)xz, \qquad (11)$$

$$dz/dt = -cz + (a-z)(x^2/b). \qquad (12)$$

The characteristic matrix of Eqs.(11) and (12) is

$$\begin{pmatrix} -v + (ka/b) - (kz/b) & -kx/b \\ (2ax/b) - (2xz/b) & -c - (x^2/b) \end{pmatrix}. \qquad (13)$$

Its characteristic equation is

$$\lambda^2 - [-v + (ka/b) - c - (kz+x^2)/b]\lambda - [-v + (ka/b)][c + (x^2/b)] + (k/b^2)(bcz + 2ax^2 - x^2z) = 0. \qquad (14)$$

A plane system of $dx/dt = 0$ and $dz/dt = 0$ has three singular points: $O(0,0)$, $A$ and $B$ ($\pm\sqrt{c(ak/v-b)}, a-(vb/k)$). For $O$ point, $\lambda_{1,2} = -v + (ka/b), -c$. When $ka/b > v$, it is a saddle point; when $ka/b < v$, it is a nodal point. For $A$ and $B$ points, $T = -(kac/vb) < 0$, so they are two stable sinks, and $\lambda_{1,2} = [kac \pm \sqrt{c(k^2a^2c - 8kabv^2 + 8b^2v^3)}]/2vb$. When $8kabv^2 > 8b^2v^3 + k^2a^2c$, they are two focal points, and the plane is a saddle-focal system;

conversely, they are two nodal points, and the plane is a saddle-nodal or nodal-nodal system.

As a particular case, when $a = 0$, $\lambda_{1,2} = \pm\sqrt{2vc}$. If $vc < 0$, A and B are two central points.

3. If $dz/dt = 0$, $z = (xy/c)$. Eqs.(3) and (4) become:

$$dx/dt = -vx + ky, \tag{15}$$

$$dy/dt = ax - by - (x^2 y)/c. \tag{16}$$

The characteristic matrix of Eqs.(15) and (16) is

$$\begin{pmatrix} -v & k \\ a - (2xy/c) & -b - (x^2/c) \end{pmatrix}. \tag{17}$$

Its characteristic equation is

$$\lambda^2 + [v + b + (x^2/c)]\lambda + v[b + (x^2/c)] - k[a - (2xy/c)] = 0. \tag{18}$$

A plane system of $dx/dt = 0$ and $dy/dt = 0$ has three singular points: $O(0,0)$, A and B ($\pm\sqrt{c(ae-b)}, \pm\sqrt{c(ae-b)}/e$). For O point, $\lambda_{1,2} = [-(v+b) \pm \sqrt{(v-b)^2 + 4ka}]/2$.

Usually, it is a saddle point or a nodal point. For A and B points, $T = -(v + ae) < 0$, so they are two stable sinks, and $\lambda_{1,2} = [-(v + ae) \pm \sqrt{(v - ae)^2 - 4ka + (8kb/e)}]/2$. When $4ka > (v - ae)^2 + (8kb/e)$, they are two focal points; conversely, they are two nodal points.

Assume that $v = 0$, Eq.(15) becomes

$$dx/dt = ky. \tag{19}$$

If Eq.(16) transforms to

$$dy/dt = (a - b)x - (x^3/c). \tag{20}$$

The characteristic roots are $\lambda_{1,2} = \pm\sqrt{k(a-b) - (3x^2/c)}$. The plane has three singular points: $O(0,0)$, A and B($\pm\sqrt{c(a-b)}, 0$). For O point, when $a > b$, it is a saddle point; when $a < b$, it is a central point. For A and B points, $\lambda_{1,2} = \pm\sqrt{(a-b)(k-3)}$. When $a > b$ and $k > 3$, the system is the structure with a saddle point and two focal points. The integral curve of Eqs.(19) and (20) is:

$$H = x^4 - 2c(a-b)x^2 + 2kcy^2. \tag{21}$$

Let $H = 0$, its range is $x^2 \le 2c(a-b), y \le \pm\sqrt{c/2k}(a-b)$. It produces a standard Roche surface, which is an evolutionary result of the binary stars.

The Lorenz model should be the middle evolutionary process through fluctuation and self-organization from chaos to order. Then nebula collapses continuously due to self-gravitation, and forms a binary star system. Probably, the jump between two "wings" in the Lorenz model describes interchanges of matter and energy between two stars.

Moreover, the Lorenz model may also be obtained from synergetics, which treats the cooperation of the parts of a complex system, and forms a macroscopic structure. The basic equations of synergetics are [21]:

$$db/dt = -kb - i\sum g_\mu \alpha_\mu + F(t),$$
$$d\alpha_\mu/dt = -\gamma\alpha_\mu + ig_\mu {}^* b\sigma_\mu + \Gamma_\mu(t), \qquad (22)$$
$$d\sigma_\mu/dt = \gamma_{11}(d_0 - \sigma_\mu) + 2i(g_\mu \alpha_\mu b^* - c.c.) + \Gamma_{\sigma,\mu}(t),$$

where b is the amplitude of field, $\alpha_\mu$ is the atomic dipole moment, and $\sigma_\mu$ is the atomic inversion. We extend synergetics to the formation of star from a nebula. In this case b is a gravitational field, $\sigma_\mu$ seems to be the density of matter. In synergetics, the quantity $\alpha_\mu$ is determined of the order parameter b [21], i.e., it obeys the gravitational field. These equations and a process of the formation of binary stars contain necessarily some stochastic effects. When we neglect various random effects $F(t), \Gamma_\mu(t), \Gamma_{\sigma,\mu}(t)$, and let $b \to x, \alpha_\mu \to y$ and $\sigma_\mu \to \sigma$, so Eqs.(22) are simplified to:

$$db/dt = dx/dt = -kx - ay, \qquad (23)$$

$$d\alpha_\mu/dt = dy/dt = -by + cx\sigma, \qquad (24)$$

$$d\sigma/dt = A - e\sigma + fxy. \qquad (25)$$

Let $z = A - e\sigma$, so Eqs.(23)(24)(25) become the equations (3)(4)(5) of the Lorenz model. This result is a space mode. When the field strength as an order parameter achieves a certain threshold, a self-organized structure is formed.

The initial energy levels of different parts in a nebula are occupied in thermal equilibrium according to the Boltzmann distribution function. When the density of matter reaches a certain threshold value, a nebula becomes unstable suddenly, and starts contraction with exponential growth under its own gravity. Probably, it is analogous to the Jeans inequality $\lambda > (\pi v_s / G\rho)^{1/2}$. It shows that a preliminary binary star system can be formed through the cooperation of different parts of a nebula. Further, two wings of the Lorenz model forms two stars, respectively, because of the gravitational force. Finally, a chaotic nebula forms a new order structure.

If the dynamic process of the order parameter in a nonlinear system is described by an equation [21]:

$$x'' + 2ax' - bx + hx^3 = 0. \qquad (26)$$

Eq.(26) may become the form:

$$\begin{cases} x' = y, \\ y' = -2ay + bx - hx^3. \end{cases} \quad (27)$$

Using the qualitative analysis theory of the equations, the characteristic matrix of Eqs.(27) is

$$\begin{pmatrix} 0 & 1 \\ b - 3hx^2 & -2a \end{pmatrix}. \quad (28)$$

Its characteristic equation is

$$\lambda^2 + 2a\lambda - b + 3hx^2 = 0, \quad (29)$$

The singular points of Eqs.(27) are $O(0,0)$, and $A_{1,2}(\pm\sqrt{b/h},0)$. For a point $O$, $T = -2b$, $D = -b < 0$, it is a saddle point. For two points $A_{1,2}$, $T_{1,2} = -2a, D_{1,2} = 2b, \Delta_{1,2} = 4a^2 - 8b$.

Because $D > 0$, $T < 0$, if $4a^2 < 8b$, they are two stable focal points; if $4a^2 > 8b$, they are two stable nodal points. The two cases all correspond to binary stars. Only when $a = 0$, i.e., the friction force may be neglected, the two points are the stable central points, which correspond to two single-stars since in this case $T_{1,2} = 0, D_{1,2} = 2b, \Delta_{1,2} = -8b$. The nonlinear force determines several singular points, but whose qualitative characters are independent of the force.

A simple example is an anharmonic oscillator, whose equation of motion is:

$$dq/dt = cq - hq^3. \quad (30)$$

For the equilibrium points $dq/dt = 0$, when $c < 0$ and $h > 0$, the only solution $q = 0$ is stable, which corresponds to the single star; when $c > 0$ and $h > 0$, we find three solutions: $q = 0$ is unstable; $q_{1,2} = \pm\sqrt{|c|/h}$ are stable, which correspond to the binary stars.

By the adiabatic approximation of synergetics, Eqs.(22) may be simplified to [21]:

$$db/dt = (-k + \frac{g^2}{\gamma}D_0)b - \frac{4g^2 k}{\gamma\gamma_\parallel}b^+bb + F(t). \quad (31)$$

If b regards as a real quantity q, Eq.(31) is evidently identical with (30), where $D = \sum_\mu \sigma_\mu$ is a total mass, $(-k + \frac{g^2}{\gamma}D_0) = c, \frac{4g^2 k}{\gamma\gamma_\parallel} = h$. When $D_0$ is bigger, we have $c > 0$. If $D_0$ is small enough, $c < 0$. The field is merely supported by stochastic processes.

The one-dimensional Fokker-Planck equation writes the form [21]:

$$f' + \frac{d}{dq}(Kf - \frac{1}{2}Q\frac{df}{dq}) = 0. \quad (32)$$

When $K(q) = cq - hq^3$, and the potential

$$V(q) = -\int_{q_0}^{q} K(q)dq = -\frac{1}{2}cq^2 + \frac{1}{4}hq^4, \tag{33}$$

its result is the same.

Further, in the plane nebula the most universal linear equations are:

$$dx/dt = ax + by, \tag{34}$$

$$dy/dt = cx - ey. \tag{35}$$

Their characteristic equation is

$$\lambda^2 - (a+e)\lambda + (ae - bc) = 0. \tag{36}$$

There is only singular point $O(0,0)$, which corresponds to a single star. When $a+e>0$, $O$ point is a unstable source; when $a+e<0$, $O$ point is a stable sink. As an example, if Eq.(1) has only a term of magnetic force, the plane equations are:

$$\begin{aligned} \frac{d\rho u_x}{dt} &= \frac{e}{c}B_z u_y \\ \frac{d\rho u_y}{dt} &= -\frac{e}{c}B_z u_x \end{aligned} \tag{37}$$

Their characteristic roots are $\lambda_{1,2} = \pm\sqrt{-(eB_z/c\rho)}$. It shows $O(0,0)$, is a central point, which corresponds to a single star. Moreover, Eqs.(1) and (3)-(5) under some particular conditions may obtain only a singular point and corresponding single star. For Eqs.(26) and (30), the nonlinear terms $h=0$, above results are the same.

Based on the nonlinear equations of hydrodynamics and synergetics, etc., we discuss the dynamical models of binary stars. In an evolutionary system the nonlinear interaction plays a crucial role, and is necessary condition of the formation of binary stars and of multiple stars, but is not sufficient condition. While various nonlinear interactions are very general, so the binary stars are common. The linear interactions form only a single star system.